\documentclass[12pt]{article}
\begin{document}
\title{CP Violation from the Neutrino Sector:\\A Case for the Superweak
Model}
\author{\normalsize B. Holdom\thanks{holdom@utcc.utoronto.ca}\\\small
{\em Department of Physics, University of Toronto}\\\small {\em Toronto,
Ontario,} M5S1A7, CANADA}\date{}\maketitle
\begin{abstract} We discuss how CP violation originating in the
right-handed neutrino sector can feed into the quark sector, in an otherwise
CP invariant theory. The dominant effects are superweak, and we suggest
that this may yield a natural resolution of the strong CP problem. This work
builds on and extends a previously proposed model of quark and lepton
masses, based on a new strong flavor interaction above the weak scale.
\end{abstract}
\baselineskip 18pt

In this paper we will propose that CP violation arises dynamically in
association with the breakdown of lepton-number, as manifested in
right-handed neutrino condensates. We will discuss how the ``leakage" of
CP violation into the quark sector can then be small, and by showing up in
4-quark operators, result in the classic superweak model of CP violation
\cite{e}. The deviations from purely real quark mass matrices may also be
small enough to naturally resolve the strong CP problem.\footnote{There
have been other proposed resolutions of the strong CP problem in the
context of the superweak model \cite{f}.} Of most immediate interest for
this picture is the prediction of the near absence of CP violation in the \(b\)
system.

Our discussion takes place in the context of dynamical symmetry breaking,
but the picture is somewhat different from a standard extended-technicolor
picture. There is a fourth family of fermions (not technifermions) whose
dynamical masses are related to electroweak symmetry breaking. There is
also a new strong flavor gauge interaction which acts on the four families
and which first breaks at a scale \(\Lambda \) in the 100 to 1000 TeV range.

When we consider the operators in the effective theory below the scale
\(\Lambda \), we find that those which can feed CP violation into the quark
sector are lepton-number violating, 6-fermion operators. For example, a
CP-violating \(\Delta S=2\) operator could be of the form
\begin{equation}{\overline{d}_{R}}{s_{L}}{\overline{d}_{R}}{s_{L}}{\nu 
_{\tau '}}{\nu _{\tau '}}.\end{equation} This is a piece of an \({{\it
SU}(2)_{L}}{\times}{{\it U}(1)_{Y}}\) invariant operator and \({\nu
_{\tau '}}\) is the heavy fourth-family left-handed neutrino. The presence of
both quarks and leptons in this operator reflects the fact that both quarks and
leptons couple to the flavor gauge interaction. If the coefficient of this
operator is of order \(1/\Lambda ^{5}\) and \({\langle }{\nu _{\tau '}}{\nu
_{\tau ' }}{\rangle }{\ \approx\ }{\Lambda _{{\it EW}}^3}\) then the
coefficient of the resulting
\({\overline{d}_{R}}{s_{L}}{\overline{d}_{R}}{s_{L}}\) operator is of
order \({\Lambda _{{\it EW}}^3}/\Lambda ^{5}\). As we will see, this can
be the appropriate size.

A theory of CP violation should also be a theory of mass, and so a
substantial fraction of this paper must be devoted to that subject. In next
section we describe the new flavor interactions and how they can give rise to
a class of operators required to generate quark and lepton masses. In section
2 we consider the CP violation in the right-handed neutrino sector and show
how it feeds into the quark sector via this same class of operators. Finally in
section 3 we describe in detail how the quark and lepton mass spectrum can
arise.

\section{Preliminaries}

A minimal flavor-gauge symmetry, \({{\it U}(2)_{V}}{\ \equiv\ }{{\it
SU}(2)_{V}}{\times}{{\it U}(1)_{V}}\), has been described in \cite{c}.
This leads to a four family model where pairs of same-charge fermions from
two of the families transform as a 2 under \({{\it U}(2)_{V}}\) and pairs
from the other two families transform as a \(\overline{2}\). We label the
quarks and leptons in these four families as \([{Q_{1}}, {L_{1}}]\),
\([{Q_{2}}, {L_{2}}]\), \([{\underline{Q}_{1}}, {\underline{L}_{1}}]\),
\([{\underline{Q}_{2}}, {\underline{L}_{2}}]\), respectively. The \(V\)
will remind us that \({{\it U}(2)_{V}}\) is a vector symmetry with respect
to these fields, which are not necessarily the mass eigenstates.

All right-handed neutrinos are assumed to have a dynamical Majorana mass
of order the flavor-physics scale \(\Lambda \). They are the only fermions to
receive mass at the flavor scale, and their condensates will serve as the order
parameters for the breakdown of \({{\it U}(2)_{V}}\) to \({{\it
U}(1)_{X}}\). If \({{\it SU}(2)_{R}}{\times}{{\it U}(1)_{B - L}}
\) is part of the weak gauge symmetry at the flavor scale, it will also be
broken in the appropriate way to \({{\it U}(1)_{Y}}\) by these neutrino
condensates. \({{\it U}(1)_{X}}\) is defined such that the \([{Q_{2}},
{L_{2}}]\) and \([{\underline{Q}_{2}}, {\underline{L}_{2}}]\) families
are \({{\it U}(1)_{X}}\) neutral (the light two families) while \([{Q_{1}},
{L_{1}}]\) and \([{\underline{Q}_{1}}, {\underline{L}_{1}}]\) have
equal and opposite \({{\it U}(1)_{X}}\) charges (the heavy two families).
\({{\it U}(1)_{X}}\) breaks close to the weak scale as we describe below,
and it should play a role in the generation of the fourth-family masses which
in turn break the electroweak symmetry.

[The fermion content of the theory and the flavor symmetry could be larger,
for example the two sets of families could transform as \({n_{f}}\) and
\({\overline{n}_{f}}\) under \({\it SU}({n_{f}})\) for \({n_{f}}{\ \geq\
}4\). In place of the breakdown \({{\it U}(2)_{V}}{\ \rightarrow\ }{{\it
U}(1)_{X}}{\ \rightarrow\ }\) nothing, we would have \({\it
SU}({n_{f}}){\ \rightarrow\ }{\it SU}({n_{f}} - 1){\ \rightarrow\ } {\it
SU}({n_{f}} - 2)\). Since we are not concerned here with trying to
understand the dynamical implications of these different choices, we will
consider \({{\it U}(2)_{V}}\) as the complete flavor symmetry for
simplicity.]

As described in \cite{c}, the dynamical fourth family masses are as follows;
the \({\it t^{\prime }}\) and \({\it b^{\prime }}\) quarks correspond to the
mass term \({\overline{\underline{Q}}_{{\it L1}}}{Q_{{\it R1}}}\) (the
hermitian conjugate term will always be implicit), the \(\tau '\) corresponds
to \({\overline{\underline{E}}_{{\it L1}}}{\underline{E}_{{\it R1}}}\),
and the left-handed \({\nu _{\tau '}}\) corresponds to
\({\underline{N}_{{\it L1}}^2}\). The \({(t^{\prime },b^{\prime })}\)
masses could be close to a TeV, while \({\nu _{\tau '}}\) may have a mass in
the few hundred GeV range, with the \(\tau '\) mass roughly twice as large.
The leptons with such masses make only small contributions (perhaps
negative) to \(S\)\cite{g} and \(T\)\cite{a}. The gauge dynamics generating
the \({(t^{\prime },b^{\prime })}\) masses is isospin symmetric, and the
small amount of \({\it t^{\prime }}\)--\({\it b^{\prime }}\) mass splitting
implied by the \(t\)--\(b\) mass splitting gives only a small contribution to
\(T\), since it is suppressed by \(({m_{t}}/{m_{{\it t^{\prime }}}})^{4}\)
\cite{c}. And finally there is the \({(t^{\prime },b^{\prime })}\)
contribution to \(S\). But since we are suggesting that the gauge dynamics
generating the \({(t^{\prime },b^{\prime })}\) masses is itself breaking
down, the theory is quite unlike QCD (i.e. there is no \(\rho \)-like
resonance), and the contribution to \(S\) is uncertain. Given all this it seems
that a fourth family with dynamical mass can still be consistent with
precision electroweak measurements.

We note that there is one additional symmetry of the flavor physics, as we
have described it. That symmetry is a \({{\it U}(1)_{A}}\) under which the
\([Q, L]\) and \([\underline{Q}, \underline{L}]\) fields have equal and
opposite \({\it axial}\) charge. Either this is a gauged symmetry which is
broken at the flavor scale, or the symmetry is already broken by 4-fermion
interactions originating at a higher scale. In either case these additional
interactions can serve to make the flavor interactions chiral, and thus
resistant to the formation of mass. 

The aspect of strong flavor dynamics crucial to our picture of quark and
lepton masses is the generation of nonperturbative multi-fermion
condensates. Given the presence of strong interactions, it is not unnatural to
expect that condensates allowed by the unbroken symmetries will form.
Their presence is especially significant when most fermions are not
receiving dynamical masses (as long as \({{\it SU}(2)_{L}}{\times}{{\it
U}(1)_{Y}}\) is an unbroken symmetry), since in that case the condensates
will imply the existence of multi-fermion operators in the effective theory
below the flavor scale.

In the presence of the fourth family masses, these operators make
contributions to the lighter quark and charged-lepton masses. When we
consider these contributions we find that the dominant contributions should
come from a particular subset of possible 4-fermion operators. Besides
being singlets under \({{\it SU}(3)_{C}}{\times}{{\it SU}(2)_{L}}
{\times}{{\it U}(1)_{Y}}\), the interesting operators have the following
properties.

\begin{itemize}
\item  They have the chiral structure \({\overline{\psi }_{L}}{\psi
_{R}}{\overline{\psi }_{L}}{\psi _{R}}\), where each \(\psi \) denotes any
quark or lepton.
\item  They preserve \({{\it SU}(2)_{V}}\) and CP.
\item  At least some or perhaps all display maximal \({{\it SU}(2)_{R}}\)
breaking.
\end{itemize}

The fact that these condensates are singlets under \({{\it SU}(2)_{V}}\)
makes dynamical sense, since it implies that they are in an attractive channel
with respect to these strong interactions. By maximal \({{\it SU}(2)_{R}}\)
breaking we mean for example that
\({\overline{Q}_{L}}{D_{R}}{\overline{\underline{Q}}_{L}}{\underline{U}_{R}}\)
is dynamically generated but not
\({\overline{Q}_{L}}{U_{R}}{\overline{\underline{Q}}_{L}}{\underline{D}_{R}}\).\footnote{One
can of course construct potentials for scalar fields where the analog of this
breaking pattern would occur for a range of parameters (see the appendix in
\cite{d}). In a similar way we could illustrate the naturalness of various
other dynamical assumptions made in this paper.} The latter can be induced
from the former, though, via an \({{\it SU}(2)_{R}}\) gauge boson
exchange; this will be our mechanism for producing the \(t\)--\(b\) mass
ratio.

Why should operators of the LRLR form dominate? One might speculate
that instanton dynamics will play a role in the generation of condensates of
the LRLR form, as opposed for example to condensates of the
\({\overline{\psi }_{L}}{\psi _{R}}{\overline{\psi }_{R}}{\psi _{L}}\)
form. Nevertheless some operators of the alternative LRRL form will be
induced by tying together a LRLR operator with the conjugate of another
LRLR operator in a loop. But even these effects may be suppressed due to
factors of \(4\pi \). The coefficients of LRLR operators are expected to be of
order \(1/f^{2}{\ \equiv\ }g^{2}/M^{2}{\ \approx\ }4\pi /M ^{2}\) where
\(M\) is the mass of a gauge boson and the strong coupling is \(g^{2}/4\pi {\
\approx\ }1\). Loop effects are then suppressed if we take \(M\) as the
ultraviolet cutoff on loop integrations and use a factor of \(1/(4\pi )^{2}\)
for each loop.

Four-fermion operators may be composed of \({{\it
SU}(2)_{V}}\)-invariant scalars like \({\overline{Q}_{{\it Li}}}{Q_{{\it
Ri}}}\) which preserve \({{\it U}(1)_{V}}\) and scalars like
\({\overline{Q}_{{\it Li}}}{\underline{Q}_{{\it Rj}}}{\varepsilon  _{{\it
ij}}}\) which do not. Four-fermion condensates which break \({{\it
U}(1)_{V}}\) will also break \({{\it U}(1)_{X}}\), and we assume that the
resulting \({m_{X}}/{g_{X}}\) is in the TeV range.\footnote{With respect
to the larger possible flavor symmetry \({\it SU}({n_{f}})\) mentioned
above, \({\overline{Q}_{{\it Li}}}{\underline{Q}_{{\it Rj}}}{\varepsilon 
_{{\it ij}}}\) with \(i, j=1, 2\) is again in an attractive channel, and
condensates containing it would break \({\it SU}({n_{f}} - 1)\) to \({\it
SU}({n_{f}} - 2)\).} The hierarchy between the \({{\it U}(2)_{V}}/{{\it
U}(1)_{X}}\) gauge boson masses and the \(X\) mass corresponds to our
expectation that contributions to gauge boson masses are larger when
coming from the 2-fermion (Majorana neutrino) condensates than when
coming from the 4-fermion condensates. The contribution from a 4-fermion
condensate involves tying the condensate together with its conjugate (three
loops), and the same loop analysis as before indicates that this is suppressed
\cite{c}.

\section{CP Violation}

Above the flavor scale we assume that we have a CP invariant gauge theory
of massless fermions. We then assume that the flavor dynamics is such that
CP violation, lepton-number violation, and \({{\it SU}(2)_{V}}\) breaking
all originate in the right-handed neutrino condensates (both bilinear and
multilinear).\footnote{The dynamical breakdown of CP naively leads to a
domain wall problem, but various resolutions of this problem have now
been proposed
\cite{i,k,l}. We learn from these references that the issue is more complex
and probably less serious than once thought.} CP violation for example
would be reflected in the phases of the Majorana mass condensates
\({\langle }{N_{{\it R2}}^2}{\rangle }\) , \({\langle }{\underline{N}_{{\it
R2}}^2}{\rangle }\), \({\langle }{N_{{\it R1}}}{\underline{N}_{{\it
R1}}} - {N_{{\it R2}} }{\underline{N}_{{\it R2}}}{\rangle }\) and
\({\langle }{N_{{\it R1}}}{\underline{N}_{{\it R1}}} + {N_{{\it R2}}
}{\underline{N}_{{\it R2}}}{\rangle }\), which are the most general
allowed by the breaking \({{\it U}(2)_{V}}{\ \rightarrow\ }{{\it
U}(1)_{X}}\). Note that only the first three break \({{\it SU}(2)_{V}}\),
and by combining a neutrino mass and a conjugate neutrino mass there are
also amplitudes which break \({{\it SU}(2)_{V}}\) and not lepton-number.
We will argue that when the right-handed neutrinos are integrated out, the
only CP-violating operators in the effective theory must violate
lepton-number or \({{\it SU}(2)_{V}}\) or both.

Let us write \({\langle }{N_{{\it R2}}^2}{\rangle }=ae^{{i\alpha }}\) ,
\({\langle }{\underline{N}_{{\it R2}}^2}{\rangle }=be^{{i\beta }}\),
\({\langle }{N_{{\it R2}}}{\underline{N}_{{\it R2}}}{\rangle }=ce
^{{i\chi }}\) where the various constants appearing here are real. Let us
consider combinations of these condensates which could appear internally
in diagrams after the right-handed neutrinos have been integrated out. Let us
first consider combinations which preserve \({N_{R}}\) number and
\({\underline{N}_{R}}\) number. Some of these combinations would be
intrinsically real, such as when a condensate and its complex conjugate
appear in a loop. When there are four condensates in a loop the phases need
not cancel, and for example one combination would be proportional to
\(abc^{2}e^{{i(\alpha  + \beta  - 2\chi )}}\). But there is another diagram in
which all condensates are replaced by their complex conjugates, and so the
sum is proportional to \(\cos(\alpha  + \beta  - 2\chi )\). The sum is thus CP
conserving, i.e. invariant under reversing the signs of all phases
simultaneously.

A similar argument applies to any combination of bilinear and multilinear
neutrino condensates. To preserve \({N_{R}}\) number and
\({\underline{N}_{R}}\) number, every neutrino line from a condensate
must be paired with an antineutrino line of the same flavor from another
condensate. Each such combination of condensates is either intrinsically
real, or when it is not there is another combination in which all condensates
are replaced by their complex conjugates so that the sum is real. Thus to find
CP violation we must consider combinations of condensates which do not
preserve \({N_{R}}\) number and/or \({\underline{N}_{R}}\) number.
These combinations produce amplitudes which break lepton-number but not
\({{\it SU}(2)_{V}}\) (such as \({N_{{\it R1}}}{\underline{N}_{{\it
R1}}} + {N_{{\it R2}}}{\underline{N}_{{\it R2}}}\)), or break \({{\it
SU}(2)_{V}}\) completely but not lepton number (such as
\({\overline{N}_{{\it R2}}}{/\!\!
\!\!D}{\underline{N}_{{\it R2}}}\)), or break both lepton-number and
\({{\it SU}(2)_{V}}\)  (such as \({N_{{\it R2}}^2}\)). Since nothing else at
the flavor scale breaks lepton number or \({{\it SU}(2)_{V}}\), the
implication is that in the effective theory below the flavor breaking scale the
only CP-violating operators are one of these three types.

We digress briefly to comment on the origin of the dynamical breakdown of
CP. We can expect a term proportional to \(\cos(\alpha  + \beta  - 2\chi )\)
(phases defined above) in some effective potential constructed to describe
the neutrino condensation. If this term has the appropriate sign, then
minimization of this one term implies that \(\alpha  + \beta  - 2\chi =\pi \).
The only CP conserving solution has the condensates real with \({\langle
}{N_{{\it R2}}^2}{\rangle }\) and \({\langle }{\underline{N}_{{\it
R2}}^2}{\rangle }\) opposite in sign. The CP-violating solutions allow
\({\langle }{N_{{\it R2}}^2}{\rangle }\) and \({\langle
}{\underline{N}_{{\it R2}}^2}{\rangle }\) to be equal but complex. Other
terms in the effective action, such as those involving multi-neutrino
condensates, can potentially pick out the latter solution.

We now consider the lepton-number and \({{\it SU}(2)_{V}}\) violating
operators in the effective theory after the right-handed neutrinos have been
integrated out. The lowest dimension \({{\it SU}(3)_{C}}{\times}{{\it
SU}(2)_{L}} {\times}{{\it U}(1)_{Y}}\) invariant operators are of
dimension 9, and interesting examples are the following.
\begin{eqnarray}&&{\overline{D}_{{\it L2}}}{D_{{\it
R2}}}{\overline{D}_{{ L2}}}{D_{{\it
R2}}}{\overline{\underline{N}}_{{\it
L1}}}{\overline{\underline{N}}_{{\it
L1}}}{\label{a}}\\&&{\overline{\underline{D}}_{{\it
L2}}}{\underline{D}_{{\it R2}}}{\overline{\underline{D}}_{{\it
L2}}}{\underline{D}_{{\it R2}}}{\overline{\underline{N}}_{{\it
L1}}}{\overline{\underline{N}}_{{\it L1}
}}{\label{b}}\\&&{\overline{D}_{{\it L2}}}{\underline{D}_{{\it
R1}}}{\overline{D} _{{\it L2}}}{\underline{D}_{{\it
R1}}}{\overline{\underline{N}}_{{\it
L1}}}{\overline{\underline{N}}_{{\it
L1}}}{\label{c}}\\&&{\overline{D}_{{\it L1}}}{\underline{D}_{{\it
R2}}}{\overline{D} _{{\it L1}}}{\underline{D}_{{\it
R2}}}{\overline{\underline{N}}_{{\it
L1}}}{\overline{\underline{N}}_{{\it L1}}}{\label{d}}\end{eqnarray}
They can be seen to arise from the \({{\it SU}(2)_{V}}\)-preserving
operators, \({\overline{D}_{{\it Li}}}{D_{{\it
Ri}}}{\overline{\underline{N}} _{{\it Lk}}}{N_{{\it Rl}}}{\varepsilon
_{{\it kl}}}\), \({\overline{\underline{D}}_{{\it
Li}}}{\underline{D}_{{\it Ri}}}{\overline{\underline{N}}_{{\it
Lk}}}{N_{{\it Rl}}}{\varepsilon _{{\it kl}}}\), and \({\overline{D}_{{\it
Li}}}{\underline{D}_{{\it Rj}}}{\varepsilon  _{{\it
ij}}}{\overline{\underline{N}}_{{\it Lk}}}{N_{{\it Rl}}} {\varepsilon
_{{\it kl}}}\), along with an insertion of the \({N_{{\it R2}}}\) mass.
\({\underline{N}_{{\it L1}}}\) is the fourth left-handed neutrino \({\nu
_{\tau '}}\) which has a Majorana mass close to the weak scale. The result
below the weak scale is an effective CP-violating 4-quark operator. In
particular the operators in (\ref{a}) and (\ref{b}) turn out to be essentially
the \(\Delta S=2\) operators \(({\overline{d}_{L}}{s_{R}})^{2}\) and
\(({\overline{s}_{L}}{d_{R}})^{2}\) respectively. This will become clear
from the quark mass matrices given in the next section.

Assuming a CP-violating phase of order unity, the coefficients of these
\(\Delta S=2\) operators (which contain pseudoscalar-pseudoscalar pieces)
should be order \(10^{{ - 10}}{\rm \ TeV^{ - 2}}\) to recover the known
value of \(\varepsilon \) in the neutral kaon system \cite{h}. If the
coefficients of the \({{\it SU}(2)_{V}}\)-preserving 4-fermion operators are
\({\ \approx\ }\Lambda ^{{ - 2}}\) and the \({\overline{N}_{{\it R2}}^2}\)
mass is \({m_{N}}\), then the coefficients of the operators in (\ref{a}) and
(\ref{b}) are \({\ \approx\ }1/(\Lambda ^{4}{m_{N}})\). It is reasonable
that this is of order \(1/(100{\rm \ TeV})^{5}\) and \({\langle
}{\underline{N}_{{\it L1}}^2}{\rangle }{\ \approx\ }(1 {\rm \
TeV})^{3}\), in which case the coefficient of the \(\Delta S=2\) operator is
of the desired size.

We have recovered the classic superweak model \cite{e} which accounts for
CP violation in \(K\)--\(\overline{K}\) mixing. CP-violating \(\Delta S=1\)
operators can be induced from those in (\ref{a}) and (\ref{b}) by mass
mixing between the \(d\) and the \(s\), but this produces a negligible
contribution to \(\varepsilon '\). \(\Delta b=2\) CP violation arises from the
operators in (\ref{c}) and (\ref{d}) which generate
\(({\overline{d}_{L}}{b_{R}})^{2}\) and
\(({\overline{b}_{L}}{d_{R}})^{2}\) operators respectively. These latter
effects would have to be \(10^{3}\) to \(10^{4}\) times larger than the
\(\Delta S=2\) CP violation in order to match the standard model prediction.
This is very unlikely, thus making the nonobservation of standard model
CP-violating effects in the \(b\) system a signature of our picture. 

Another possible signal of CP violation in the quark sector is in the QCD
vacuum angle \(\overline{\theta }\). In the underlying CP-invariant theory
of massless fermions the QCD vacuum angle \(\theta \) vanishes, but a
nonzero \(\overline{\theta }\) can be generated if CP violation in the
neutrino sector feeds into the quark mass matrix. In fact a possibly
dangerous contribution arises if the operator
\({\overline{\underline{E}}_{{\it L1}}}{\underline{E}_{{\it
R1}}}{\overline{\underline{N}}_{{\it L1}}}{N_{{\it R2}}}\) is
dynamically generated. Along with the operators considered above it would
generate the following 6-fermion operators.
\begin{eqnarray}&&{\overline{D}_{{\it L2}}}{D_{{\it
R2}}}{\overline{\underline{E}} _{{\it L1}}}{\underline{E}_{{\it
R1}}}{\overline{\underline{N}}_{{\it
L1}}}{\overline{\underline{N}}_{{\it
L1}}}\\&&{\overline{\underline{D}}_{{\it L2}}}{\underline{D}_{{\it
R2}}}{\overline{\underline{E}}_{{\it L1}}}{\underline{E}_{{\it
R1}}}{\overline{\underline{N}}_{{\it
L1}}}{\overline{\underline{N}}_{{\it L1} }}\\&&{\overline{D}_{{\it
Li}}}{\underline{D}_{{\it Rj}}}{\varepsilon  _{{\it
ij}}}{\overline{\underline{E}}_{{\it L1}}}{\underline{E}_{{\it
R1}}}{\overline{\underline{N}}_{{\it
L1}}}{\overline{\underline{N}}_{{\it L1}}}\end{eqnarray}
 \({\overline{\underline{E}}_{{\it L1}}}{\underline{E}_{{\it R1}}}\)
corresponds to the \(\tau '\) mass, and so along with the \({\nu _{\tau '}}\)
mass these operators could make CP-violating contributions to the
\({\overline{d}_{L}}{s_{R}}\), \({\overline{s}_{L}}{d_{R}}\),
\({\overline{d}_{L}}{b_{R}}\) or \({\overline{b}_{L}}{d_{R}}\)
off-diagonal mass terms. If the coefficients of these operators are of order
\(1/(100{\rm \ TeV})^{5}\) and \({\langle }{\overline{\underline{E}}_{{\it
L1}}}{\underline{E}_{{\it R1 }}}{\rangle }{\ \approx\ }(1{\rm \
TeV})^{3}\) then the contribution to the imaginary parts of these mass
terms could be as large as roughly 100 eV. By comparing to light quark
masses we see that the suppression arising from the small size of generic
6-fermion operators may not sufficiently suppress \(\overline{\theta
}\).\footnote{The current experimental upper bound on the neutron EDM is
satisfied for \(\overline{\theta }{\ \approx\ }10^{{ - 9}}\) \cite{j}.}

The largest contribution to \(\overline{\theta }\) will likely come from the
\(d\)--\(s\) mass elements. Given that the diagonal elements of the down
quark mass matrix dominate the determinant (see next section), we have
\begin{equation}\overline{\theta }{\ \approx\ } - \frac {{\rm Re} ({m_{{\it
ds}}}){\rm Im} ({m_{{\it sd}}}) + {\rm Re} ({m_{{\it sd}}}){\rm Im}
({m_{{\it ds}}})}{{m_{d}}{m_{s}}}.\end{equation} We can now
identify additional possible sources of suppression which make an
acceptable value for \(\overline{\theta }\) fairly plausible.

\begin{itemize}
\item  In the next section we will see that \({\rm Re} ({m_{{\it sd}}})\) and
\({\rm Re} ({m_{{\it ds}}})\) are suppressed because they can only be
generated by 4-fermion operators of the suppressed LRRL form. (In the
up-sector on the other hand, LRLR operators contribute to the off-diagonal
terms, which could then be the origin of most of the Cabbibo mixing.)
\item  The offending \({\overline{\underline{E}}_{{\it
L1}}}{\underline{E}_{{\it R1}}}{\overline{\underline{N}}_{{\it
L1}}}{N_{{\it R2}}}\) operator may be one of the operators disfavored
due to the maximal breakdown of \({{\it SU}(2)_{R}}\), in which case it
may only arise as a radiative correction to the operator
\({\overline{\underline{E}}_{{\it L1}}}{\underline{N}_{{\it
R1}}}{\overline{\underline{N}}_{{\it L1}}}{E_{{\it R2}}}\). These two
operators would then be analogous to the \({\tilde{\cal{D}}}\) and
\({\cal{D}}\) 4-quark operators appearing in the next section.
\item  As for CP violation leaking into the up-sector masses, in addition
to\linebreak \({\overline{\underline{E}}_{{\it L1}}}{\underline{E}_{{\it
R1}}}{\overline{\underline{N}}_{{\it L1}}}{N_{{\it R2}}}\) we would
need operators like \({\overline{U}_{{\it R2}}}{U_{{\it
L2}}}{\overline{\underline{N}} _{{\it L1}}}{N_{{\it R2}}}\) which are
of the suppressed LRRL form. The generation of such operators may be
further suppressed due to the maximal breakdown of \({{\it SU}(2)_{R}}\).
\item  Due to the absence of color interactions it is conceivable that purely
leptonic operators (or at least those which break \({{\it U}(1)_{V}}\) and
\({{\it U}(1)_{X}}\)) are generated only through loops (at least two)
involving other LRLR operators. In fact we will see that purely leptonic
operators of dynamical origin are not required for the generation of quark or
charged-lepton masses. 
\end{itemize}

\section{Quark and Lepton Masses}

We first describe the quark masses in a manner similar to, but not identical
to, a previous description \cite{c}. We then turn to a description of lepton
masses which is essentially new. We will see that quarks and charged-lepton
masses may be completely described in terms of operators of the LRLR
form. We will also highlight the interplay between the quark and lepton
sectors.

We first consider 4-quark operators. In the following list we have labeled
those pieces of \({{\it SU}(2)_{V}}\)-invariant operators which make
important contributions to the quark masses. Only the \({\cal{B}}\) and
\({\tilde{\cal{B}}}\) operators preserve both \({{\it U}(1)_{V}}\) and
\({{\it U}(1)_{A}}\).
\begin{equation}
 \left[  {\begin{array}{cc} {\overline{U}_{{\it L1}}}{D_{{\it
R1}}}{\overline{\underline{D}} _{{\it L1}}}{\underline{U}_{{\it R1}}}
& {\cal{B}} \\ {\overline{D}_{{\it L1}}}{U_{{\it
R1}}}{\overline{\underline{U}} _{{\it L1}}}{\underline{D}_{{\it R1}}}
& {\tilde{\cal{B}}} \\ {\overline{U}_{{\it L1}}}{D_{{\it
R1}}}{\overline{\underline{D}} _{{\it L1}}}{U_{{\it R2}}} & {\cal{C}}
\\ {\overline{D}_{{\it L1}}}{U_{{\it R1}}}{\overline{\underline{U}}
_{{\it L1}}}{D_{{\it R2}}} & {\tilde{\cal{C}}} \\
{\overline{\underline{U}}_{{\it L2}}}{D_{{\it
R1}}}{\overline{\underline{D}}_{{\it L1}}}{\underline{U}_{{\it R1}}}
& {\cal{D}} \\ {\overline{\underline{D}}_{{\it L2}}}{U_{{\it
R1}}}{\overline{\underline{U}}_{{\it L1}}}{\underline{D}_{{\it R1}}}
& {\tilde{\cal{D}}} \\ {\overline{\underline{Q}}_{{\it Li}}}{U_{{\it
Rj}}}{\varepsilon  _{{\it ij}}}{\overline{\underline{Q}}_{{\it
Lk}}}{D_{{\it Rl}}} {\varepsilon _{{\it kl}}} & {\cal{E}} \\
{\overline{Q}_{{\it Li}}}{\underline{U}_{{\it Rj}}}{\varepsilon  _{{\it
ij}}}{\overline{Q}_{{\it Lk}}}{\underline{D}_{{\it Rl}}} {\varepsilon
_{{\it kl}}} & {\cal{F}}
\end{array}}
 \right] 
\end{equation} These operators feed mass down to the known three families
of quarks from the \({\it t^{\prime }}\) and \({\it b^{\prime }}\) masses
(\({\overline{\underline{U}}_{{\it L1}}}{U_{{\it R1}}}\) and
\({\overline{\underline{D}}_{{\it L1}}}{D_{{\it R1}}}\)) except for the
\({\cal{F}}\) operator, which feeds mass down from the \(t\) mass
(\({\overline{U}_{{\it L1}}}{\underline{U}_{{\it R1}}}\)). The \({\it
t^{\prime }}\) and \({\it b^{\prime }}\) masses have to be close to
degenerate and so the \(t\)--\(b\) mass ratio must be due to \({{\it
SU}(2)_{R}}\) breaking in these operators. If there is a dynamical
breakdown of \({{\it SU}(2)_{R}}\) then we could suppose that the
\({\cal{B}}\), \({\cal{C}}\) and \({\cal{D}}\) operators are generated but
not the \({\tilde{\cal{B}}}\), \({\tilde{\cal{C}}}\) and \({\tilde{\cal{D}}}\)
operators. If \({{\it SU}(2)_{R}}\) is a weak gauge symmetry at the flavor
scale then the latter operators will be induced from the former operators by
an \({{\it SU}(2)_{R}}\) gauge boson exchange. In this way the \(b\) mass
arises as a radiative correction to the \(t\) mass.

Important contributions to the quark masses will also feed in from the lepton
sector. The following mixed quark-lepton operators feed mass down from
the \(\tau '\) mass (\({\overline{\underline{E}}_{{\it
L1}}}{\underline{E}_{{\it R1}}}\)). Only the \({\cal{G}}\) operators
preserve both \({{\it U}(1)_{V}}\) and \({{\it U}(1)_{A}}\).
\begin{equation}
 \left[  {\begin{array}{cc} {\overline{\underline{E}}_{{\it
L1}}}{\underline{E}_{{\it R1}}}{\overline{U}_{{\it L1}}}{U_{{\it
R1}}} & {{\cal{G}}_{1}} \\ {\overline{\underline{E}}_{{\it
L1}}}{\underline{E}_{{\it R1}}}{\overline{U}_{{\it L2}}}{U_{{\it
R2}}} & {{\cal{G}}_{2}} \\ {\overline{\underline{E}}_{{\it
L1}}}{\underline{E}_{{\it R1}}}{\overline{\underline{U}}_{{\it
L1}}}{\underline{U}_{{\it R1}}} & {{\cal{H}}_{1}} \\
{\overline{\underline{E}}_{{\it L1}}}{\underline{E}_{{\it
R1}}}{\overline{\underline{U}}_{{\it L2}}}{\underline{U}_{{\it R2}}}
& {{\cal{H}}_{2}} \\ {\overline{\underline{E}}_{{\it
L1}}}{\underline{E}_{{\it R1}}}{\overline{U}_{{\it
Li}}}{\underline{U}_{{\it Rj}}}{\varepsilon _{{\it ij}}} & {\cal{I}} \\
{\overline{\underline{E}}_{{\it L1}}}{\underline{E}_{{\it
R1}}}{\overline{\underline{U}}_{{\it Li}}}{U_{{\it Rj}}}{\varepsilon
_{{\it ij}}} & {\cal{J}}
\end{array}}
 \right] 
\end{equation} We write the quark mass matrices in terms of the original
fields as follows, where the \({\it t^{\prime }}\) and \({\it b^{\prime }}\)
masses correspond to the bottom right corner.
\begin{equation}
 \left[  {\begin{array}{cccc} {\overline{Q}_{{\it
L2}}}{\underline{Q}_{{\it R2}}} & {\overline{Q} _{{\it L2}}}{Q_{{\it
R2}}} & {\overline{Q}_{{\it L2}}}{\underline{Q}_{{\it R1}}} &
{\overline{Q}_{{\it L2}}}{Q_{{\it R1}}} \\
{\overline{\underline{Q}}_{{\it L2}}}{\underline{Q}_{{\it R2}}} &
{\overline{\underline{Q}}_{{\it L2}}}{Q_{{\it R2}}} &
{\overline{\underline{Q}}_{{\it L2}}}{\underline{Q}_{{\it R1}}} &
{\overline{\underline{Q}}_{{\it L2}}}{Q_{{\it R1}}} \\
{\overline{Q}_{{\it L1}}}{\underline{Q}_{{\it R2}}} & {\overline{Q}
_{{\it L1}}}{Q_{{\it R2}}} & {\overline{Q}_{{\it
L1}}}{\underline{Q}_{{\it R1}}} & {\overline{Q}_{{\it L1}}}{Q_{{\it
R1}}} \\ {\overline{\underline{Q}}_{{\it L1}}}{\underline{Q}_{{\it
R2}}} & {\overline{\underline{Q}}_{{\it L1}}}{Q_{{\it R2}}} &
{\overline{\underline{Q}}_{{\it L1}}}{\underline{Q}_{{\it R1}}} &
{\overline{\underline{Q}}_{{\it L1}}}{Q_{{\it R1}}}
\end{array}}
 \right] 
\end{equation} Here then are the contributions from the various operators.
\begin{equation} {M_{u}}= \left[  {\begin{array}{cccc} 0 &
{{\cal{G}}_{2}} & {\cal{I}} & 0 \\ {{\cal{H}}_{2}} & {\cal{E}} &
{\cal{D}} & {\cal{J}} \\ {\cal{I}} & {\cal{C}} & {\cal{B}} &
{{\cal{G}}_{1}} \\ 0 & {\cal{J}} & {{\cal{H}}_{1}} & {\cal{A}}
\end{array}}
 \right] 
\end{equation}
\begin{equation} {M_{d}}= \left[  {\begin{array}{cccc} {\cal{F}} & 0 &
0 & 0 \\ 0 & {\cal{E}} & {\tilde{\cal{D}}} & 0 \\ 0 & {\tilde{\cal{C}}} &
{\tilde{\cal{B}}} & 0 \\ 0 & 0 & 0 & {\cal{A}}
\end{array}}
 \right] 
\end{equation} None of the zero entries are exactly zero; in \({M_{u}}\)
these entries are too small to have any significance while in \({M_{d}}\)
some could be significant, but they must be generated by operators of the
suppressed LRRL form.

The following points are relevant to understanding the various hierarchies. 

\begin{itemize}
\item  The operators have different transformation properties under the
strong \({{\it U}(1)_{X}}\), and this will cause different anomalous
power-law scaling enhancements as the operators are run down from the
flavor scale to a TeV.
\begin{eqnarray}&&{\cal{B}}>{\cal{C}},
{\cal{D}}>{\cal{E}}\\&&{{\cal{G}}_{1}}, {{\cal{H}}_{1}}>{\cal{I}},
{\cal{J}}> {{\cal{G}}_{2}}, {{\cal{H}}_{2}}\end{eqnarray}
\item  There are different heavy masses, \({m_{{\it t^{\prime },b^{\prime
}}}}>{m_{\tau '}} >{m_{t}}\), being fed down.
\begin{eqnarray}&&{\cal{E}}>{\cal{F}}\\&&{\cal{B}}>{{\cal{G}}_{1}},
{{\cal{H}}_{1}}\\&&{\cal{C}}, {\cal{D}}>{\cal{I}},
{\cal{J}}\end{eqnarray}
\item  \({\tilde{\cal{B}}}\), \({\tilde{\cal{C}}}\) and \({\tilde{\cal{D}}}\)
arise from weak radiative corrections.
\begin{equation}{\cal{B}}, {\cal{C}}, {\cal{D}}>{\tilde{\cal{B}}},
{\tilde{\cal{C}}}, {\tilde{\cal{D}}}\end{equation}
\item  Some operators break \({{\it U}(1)_{A}}\) while others do not. Thus,
for example,
\begin{equation}{\cal{G}}>{\cal{H}}.\end{equation}
\end{itemize}

We note that the \({\cal{E}}\) entry is the same in the two matrices, since
that operator is intrinsically \({{\it SU}(2)_{R}}\) conserving. If this entry
determines the \(s\) mass then the \({\cal{C}}\) and \({\cal{D}}\) entries
must be responsible for the \(c\) mass, by causing mixing with the \(t\). 
Similar in size to the \({\cal{E}}\) operator is the \({\cal{F}}\) operator,
which feeds mass from the \(t\) to the \(d\). We thus expect that
\begin{equation}\frac {{m_{d}}}{{m_{s}}}{\ \approx\ }\frac
{{m_{t}}}{{m_{{\it t^{\prime }}}}}.\end{equation}

Examples of matrices which give realistic masses\footnote{The up-type
masses are (.002,.74, 160, 1000) GeV and the down-type masses are
basically the diagonal entries; these values are appropriate for masses
renormalized at a TeV.} and mixings are the following.
\begin{equation} {M_{u}}= \left[  {\begin{array}{cccr} 0 & .1 & 1 & 0 \\
-.025 & .1 & 10 & 1 \\ -1 & -10 & 160 & 10 \\ 0 & -1 & -2.5 & 1000
\end{array}}
 \right] 
\end{equation}
\begin{equation} {M_{d}}= \left[  {\begin{array}{cccr} .005 & 0 & 0 & 0
\\ 0 & .1 & .07 & 0 \\ 0 & -.07 & 3 & 0 \\ 0 & 0 & 0 & 1000
\end{array}}
 \right] 
\end{equation}

We now turn to the charged-lepton masses, where the mixed quark-lepton
operators again play an essential role. The following operators will feed
mass down from the \({\it t^{\prime }}\),
\begin{equation}
 \left[  {\begin{array}{cc} {\overline{E}_{{\it L1}}}{U_{{\it
R1}}}{\overline{\underline{U}} _{{\it L1}}}{\underline{E}_{{\it R1}}}
& {{\cal{B}}_{{\ell}}} \\ {\overline{E}_{{\it L1}}}{U_{{\it
R1}}}{\overline{\underline{U}} _{{\it L1}}}{E_{{\it R2}}} &
{{\cal{C}}_{{\ell}} } \\ {\overline{\underline{E}}_{{\it L2}}}{U_{{\it
R1}}}{\overline{\underline{U}}_{{\it L1}}}{\underline{E}_{{\it R1}}}
& {{\cal{D}}_{{\ell}}} \\ {\overline{\underline{E}}_{{\it L2}}}{U_{{\it
R1}}}{\overline{\underline{U}}_{{\it L1}}}{E_{{\it R2}}} &
{{\cal{E}}_{{\ell}}}
\end{array}}
 \right] 
\end{equation} while the following operators will feed mass down from the
\(t\).
\begin{equation}
 \left[  {\begin{array}{cc} {\overline{\underline{E}}_{{\it
L1}}}{\underline{U}_{{\it R1}}}{\overline{U}_{{\it L1}}}{E_{{\it
R1}}} & {{\cal{F}}_{{\ell}}} \\ {\overline{E}_{{\it
L2}}}{\underline{U}_{{\it R1}}}{\overline{U} _{{\it L1}}}{E_{{\it
R1}}} & {{\cal{G}}_{{\ell}} } \\ {\overline{\underline{E}}_{{\it
L1}}}{\underline{U}_{{\it R1}}}{\overline{U}_{{\it
L1}}}{\underline{E}_{{\it R2}}} & {{\cal{H}}_{{\ell}}} \\
{\overline{E}_{{\it L2}}}{\underline{U}_{{\it R1}}}{\overline{U} _{{\it
L1}}}{\underline{E}_{{\it R2}}} & {{\cal{I}}_{{\ell}}}
\end{array}}
 \right] 
\end{equation} Only the \({{\cal{B}}_{{\ell}}}\) and
\({{\cal{F}}_{{\ell}}}\) operators preserve both \({{\it U}(1)_{V}}\) and
\({{\it U}(1)_{A}}\). There are also purely leptonic operators of interest
which, unlike all the other operators we have considered in this paper, are
generated by the exchange of massive \({{\it SU}(2)_{V}}\) gauge bosons.
We will label two operators of this type, \({\overline{\underline{E}}_{{\it
L1}}}{\underline{E}_{{\it R1}}}{\overline{\underline{E}}_{{\it
R2}}}{\underline{E}_{{\it L2}}}\) and \({\overline{E}_{{\it
L1}}}{E_{{\it R1}}}{\overline{E}_{{ R2}}}{E_{{\it L2}}}\), by
\({{\cal{J}}_{{\ell}}}\) and \({{\cal{K}}_{{\ell}}}\).

We write the charged-lepton mass matrix as follows, where the large \(\tau
'\) mass is in the bottom right corner.
\begin{equation}
 \left[  {\begin{array}{cccc} {\overline{E}_{{\it L2}}}{E_{{\it R2}}} &
{\overline{E}_{{ L2}}}{\underline{E}_{{\it R2}}} & {\overline{E}_{{\it
L2}}}{E_{{\it R1}}} & {\overline{E}_{{\it L2}}}{\underline{E}_{{\it
R1}}} \\ {\overline{\underline{E}}_{{\it L2}}}{E_{{\it R2}}} &
{\overline{\underline{E}}_{{\it L2}}}{\underline{E}_{{\it R2}}} &
{\overline{\underline{E}}_{{\it L2}}}{E_{{\it R1}}} &
{\overline{\underline{E}}_{{ L2}}}{\underline{E}_{{\it R1}}} \\
{\overline{E}_{{\it L1}}}{E_{{\it R2}}} & {\overline{E}_{{
L1}}}{\underline{E}_{{\it R2}}} & {\overline{E}_{{\it L1}}}{E_{{\it
R1}}} & {\overline{E}_{{\it L1}}}{\underline{E}_{{\it R1}}} \\
{\overline{\underline{E}}_{{\it L1}}}{E_{{\it R2}}} &
{\overline{\underline{E}}_{{\it L1}}}{\underline{E}_{{\it R2}}} &
{\overline{\underline{E}}_{{\it L1}}}{E_{{\it R1}}} &
{\overline{\underline{E}}_{{ L1}}}{\underline{E}_{{\it R1}}}
\end{array}}
 \right] 
\end{equation} The various operators contribute as follows.
\begin{equation} {M_{{\ell}}}= \left[  {\begin{array}{cccc}
{{\cal{K}}_{{\ell}}} & {{\cal{I}}_{{\ell}}} & {{\cal{G}}_{{\ell}}} & 0
\\ {{\cal{E}}_{{\ell}}} & {{\cal{J}}_{{\ell}}} & 0 &
{{\cal{D}}_{{\ell}}} \\ {{\cal{C}}_{{\ell}}} & 0 & 0 &
{{\cal{B}}_{{\ell}}} \\ 0 & {{\cal{H}}_{{\ell}}} & {{\cal{F}}_{{\ell}}}
& {{\cal{A}}_{{\ell}}}
\end{array}}
 \right] 
\end{equation}

We see that the \({{\cal{B}}_{{\ell}}}\) and \({{\cal{F}}_{{\ell}}}\)
operators are essential for the generation of the \(\tau \) mass, and we note
that these operators are the analog of the dominant \({\cal{B}}\) operator in
the quark sector which generated the \(t\) mass. The
\({{\cal{K}}_{{\ell}}}\) operator then feeds the resulting \(\tau \) mass
down to the electron mass. It seems reasonable for the
\({{\cal{J}}_{{\ell}}}\) operator to give the \(\mu \) mass, since its
coefficient would have to be \({\ \approx\ }1/(100{\rm \ TeV})^{2}\)
assuming that \({\langle }{\overline{\underline{E}}_{{\it
L1}}}{\underline{E}_{{\it R1 }}}{\rangle }{\ \approx\ }(1{\rm \
TeV})^{3}\). If in fact the \({{\cal{J}}_{{\ell}}}\) and
\({{\cal{K}}_{{\ell}}}\) are the dominant contributions to the \(\mu \) and
\(e\) masses then we expect that
\begin{equation}\frac {{m_{e}}}{{m_{\mu }}}{\ \approx\ }\frac
{{m_{\tau }}}{{m_{\tau '}}}.\end{equation} The remaining zeros in the
charged-lepton mass matrix would be filled in by operators of the
suppressed LRRL form.

Remaining to be discussed are the three light left-handed neutrinos, \({\nu
_{e}}\), \({\nu _{\mu }}\), \({\nu _{\tau }}\). Their Majorana masses are
generated from 6-fermion operators, which leads to a natural suppression of
these masses compared to all other masses. Such operators are generated
from purely leptonic \({{\it SU}(2)_{V}}\)-invariant 4-fermion operators;
for example two \({\overline{\underline{E}}_{{\it
L1}}}{\underline{E}_{{\it R1}}}{\overline{N}_{{\it L2}}}{N_{{\it
R2}}}\) operators along with the large \({N_{{\it R2}}}\) mass can produce
the operator
\begin{equation}{\overline{\underline{E}}_{{\it
L1}}}{\underline{E}_{{\it R1}}}{\overline{\underline{E}}_{{\it
L1}}}{\underline{E}_{{\it R1}}}{\overline{N}_{{\it
L2}}}{\overline{N}_{{\it L2}}}.\end{equation} This, in the presence of
the \(\tau '\) mass, produces a small \({N_{{\it L2}}}\) (i.e. \({\nu _{e}}\))
mass. This neutrino mass is naively of the same order (\({\ \approx\ }100\)
eV) as the CP-violating contributions to the quark masses, although some of
the additional sources of suppression mentioned there can also apply here.

We will summarize the possible combinations of 4-fermion operators and
right-handed neutrino masses which produce left-handed neutrino masses.
We write the left-handed neutrino mass matrix as follows.
\begin{equation}
 \left[  {\begin{array}{cccc} {N_{{\it L2}}^2} & {N_{{\it
L2}}}{\underline{N}_{{\it L2}}} & {N _{{\it L2}}}{N_{{\it L1}}} &
{N_{{\it L2}}}{\underline{N}_{{ L1}}} \\ {N_{{\it
L2}}}{\underline{N}_{{\it L2}}} & {\underline{N}_{{\it L2}}^2} &
{\underline{N}_{{\it L2}}}{N_{{\it L1}}} & {\underline{N}_{{\it L2
}}}{\underline{N}_{{\it L1}}} \\ {N_{{\it L2}}}{N_{{\it L1}}} &
{\underline{N}_{{\it L2}}}{N_{{\it L1}}} & {N_{{\it L1}}^2} &
{N_{{\it L1}}}{\underline{N}_{{\it L1}}} \\ {N_{{\it
L2}}}{\underline{N}_{{\it L1}}} & {\underline{N}_{{\it L2}}}
{\underline{N}_{{\it L1}}} & {N_{{\it L1}}}{\underline{N}_{{\it L1}}}
 & {\underline{N}_{{\it L1}}^2}
\end{array}}
 \right] 
\end{equation} The large \({\nu _{\tau '}}\) mass in the bottom right corner
essentially decouples from the rest, and so we will just consider the
operators relevant to the remaining \({3\times\ 3}\) matrix.
\begin{equation}
 \left[  {\begin{array}{cc} {\overline{\underline{E}}_{{\it
L1}}}{\underline{E}_{{\it R1}}}{\overline{N}_{{\it L2}}}{N_{{\it
R2}}} & {{\cal{B}}_{\nu }} \\ {\overline{\underline{E}}_{{\it
L1}}}{\underline{E}_{{\it R1}}}{\overline{\underline{N}}_{{\it
L2}}}{\underline{N}_{{\it R2}}} & {{\cal{C}}_{\nu }} \\
{\overline{\underline{E}}_{{\it L1}}}{\underline{E}_{{\it
R1}}}{\overline{N}_{{\it L1}}}{\underline{N}_{{\it R2}}} &
{{\cal{D}}_{\nu }} \\ {\overline{\underline{E}}_{{\it
L1}}}{\underline{E}_{{\it R1}}}{\overline{N}_{{\it L1}}}{N_{{\it
R1}}} & {{\cal{E}}_{\nu }} \\ {\overline{\underline{E}}_{{\it
L1}}}{\underline{E}_{{\it R1}}}{\overline{\underline{N}}_{{\it
L2}}}{N_{{\it R1}}} & {{\cal{F}}_{\nu }} \\
{\overline{\underline{E}}_{{\it L1}}}{\underline{E}_{{\it
R1}}}{\overline{N}_{{\it L2}}}{\underline{N}_{{\it R1}}} &
{{\cal{G}}_{\nu }}
\end{array}}
 \right] 
\end{equation} We label the right-handed neutrino masses as follows.
\begin{equation}
 \left[  {\begin{array}{cc} {N_{{\it R2}}^2} & {m_{1}} \\
{\underline{N}_{{\it R2}}^2} & {m_{2}} \\ {N_{{\it
R2}}}{\underline{N}_{{\it R2}}} & {m_{3}} \\ {N_{{\it
R1}}}{\underline{N}_{{\it R1}}} & {m_{4}}
\end{array}}
 \right] 
\end{equation} The left-handed masses then arise from the following
combinations of operators and right-handed neutrino masses.
\begin{equation}
 \left[  {\begin{array}{ccc} {\displaystyle \frac {{{\cal{B}}_{\nu
}^2}}{{m_{1}}}}  &  {\displaystyle \frac {{{\cal{B}}_{\nu
}}{{\cal{C}}_{\nu }}}{{m _{3}}}}  + {\displaystyle \frac
{{{\cal{F}}_{\nu }}{{\cal{G}}_{\nu }}}{{m_{4}}}}  & {\displaystyle
\frac {{{\cal{B}}_{\nu }}{{\cal{D}}_{\nu }}}{{m_{3}}}}  +
{\displaystyle \frac {{{\cal{E}} _{\nu }}{{\cal{G}}_{\nu }}}{{m_{4}}}} 
\\ [2ex] {\displaystyle \frac {{{\cal{B}}_{\nu }}{{\cal{C}}_{\nu }}}{{m
_{3}}}}  + {\displaystyle \frac {{{\cal{F}}_{\nu }}{{\cal{G}}_{\nu
}}}{{m_{4}}}}  & {\displaystyle \frac {{{\cal{C}}_{\nu }^2}
}{{m_{2}}}}  & {\displaystyle \frac {{{\cal{C}}_{\nu }}{{\cal{D}}_{\nu
}}}{{m_{2}}}}  \\ [2ex] {\displaystyle \frac {{{\cal{B}}_{\nu
}}{{\cal{D}}_{\nu }}}{{m _{3}}}}  + {\displaystyle \frac
{{{\cal{E}}_{\nu }}{{\cal{G}}_{\nu }}}{{m_{4}}}}  & {\displaystyle
\frac {{{\cal{C}}_{\nu }}{{\cal{D}}_{\nu }}}{{m_{2}}}}  &
{\displaystyle \frac {{{\cal{D}} _{\nu }^2}}{{m_{2}}}} 
\end{array}}
 \right] 
\end{equation} This matrix can take a very different form from the quark
and charged-lepton mass matrices. For example it would not be unnatural to
assume that the masses \({m_{1}}\), \({m_{2}}\), \({m_{3}}\),
\({m_{4}}\) are similar, and that \({{\cal{B}}_{\nu }}\) and
\({{\cal{C}}_{\nu }}\) are similar, in which case \({\nu _{e}}\) and \({\nu
_{\mu }}\) could have similar mass and enjoy large mixing.
\({{\cal{D}}_{\nu }}\) could be smaller than \({{\cal{B}}_{\nu }}\) and
\({{\cal{C}}_{\nu }}\), in which case \({\nu _{\tau }}\) could be lighter
than \({\nu _{e}}\) and \({\nu _{\mu }}\). We also note that the
\({{\cal{E}}_{\nu }}\) operator, which contributes to \({\nu _{e}}\)--\({\nu
_{\tau }}\) mixing, enjoys the most enhancement from \({{\it U}(1)_{X}}\)
scaling. And finally we note that since the right-handed neutrino masses are
involved in generating this matrix, large CP-violating phases can be present.

We now briefly discuss other flavor-changing effects, all of which appear to
be at suitably suppressed levels. 

\begin{itemize}
\item  We have mentioned above that CP violation could also show up in
lepton-number conserving, \({{\it SU}(2)_{V}}\)-violating operators. For
example a \({N_{R}}\) to \({\underline{N}_{R}}\) transition inside a loop
involving a \({{\it W}_{R}}\) could induce a \(\mu \)--\(e\)--\(\gamma \)
coupling, which along with \(\mu \)--\(e\) mass mixing could generate
electron and muon electric dipole moments. Even ignoring \(\mu \)--\(e\)
mass mixing suppression, the moments are sufficiently suppressed by the
large masses of right-handed neutrinos and \({{\it W}_{R}}\). The decay
\(\mu {\ \rightarrow\ }e\gamma \), as well as \(\mu {\ \rightarrow\ }3e\) and
\(\mu \)--\(e\) conversion from the \(\mu \)--\(e\)--\(Z\) coupling, are well
below current bounds for the same reason.
\item  \(K\)--\(\overline{K}\) mixing could arise from the operator
\({\overline{D}_{{\it L2}}}{D_{{\it R2}}}{\overline{\underline{D}}
_{{\it R2}}}{\underline{D}_{{\it L2}}}\) which corresponds to
\({\overline{d}_{L}}{s_{R}}{\overline{d}_{R}}{s_{L}}\). \(K{\
\rightarrow\ }e^{-}\mu ^{+}\) could arise from the operators
\({\overline{E}_{{\it L2}}}{D_{{\it R2}}}{\overline{\underline{D}}
_{{\it R2}}}{\underline{E}_{{\it L2}}}\) and \({\overline{E}_{{\it
R2}}}{D_{{\it L2}}}{\overline{\underline{D}} _{{\it
L2}}}{\underline{E}_{{\it R2}}}\), which correspond to 
\({\overline{e}_{L}}{s_{R}}{\overline{d}_{R}}{\mu _{L}}\) and
\({\overline{e}_{R}}{d_{L}}{\overline{s}_{L}}{\mu _{R}}\). All these
operators break \({{\it U}(1)_{A}}\) and are of the suppressed LRRL form.
They also receive no enhancement from \({{\it U}(1)_{X}}\) scaling.
\item  The exchange of an \({{\it U}(2)_{V}}\) gauge boson produces the
\({\overline{s}_{L}}{s_{R}}{\overline{s}_{R}}{s_{L}}\) and
\({\overline{\mu }_{L}}{s_{R}}{\overline{s}_{R}}{\mu _{L}}\)
operators for example, which can give rise to \(K\)--\(\overline{K}\) mixing
and \(K{\ \rightarrow\ }e^{-}\mu ^{+}\) in the presence of appropriate mass
mixing in the down-quark and charged-lepton sectors. But we have seen
how mass mixings in these sectors are suppressed. Since there is more mass
mixing in the up-quark sector the corresponding effects for
\(D\)--\(\overline{D}\) mixing should be somewhat larger.
\item  \({B_{d}}\)--\({\overline{B}_{d}}\) mixing could arise from the
operator \({\overline{D}_{{\it L1}}}{\underline{D}_{{\it
R2}}}{\overline{\underline{D}}_{{\it R1}}}{D_{{\it L2}}}\) which
corresponds to
\({\overline{b}_{L}}{d_{R}}{\overline{b}_{R}}{d_{L}}\). This is again
of the suppressed LRRL form, although it would be \({{\it U}(1)_{X}}\)
enhanced. Lastly, \(X\) gauge boson exchange can give rise to
\({B_{d}}\)--\({\overline{B}_{d}}\), \({B_{s}}\)--\({\overline{B}_{s}}\)
and \(D\)--\(\overline{D}\) mixing given the appropriate mass mixings
(which are suppressed for the \(b\)). 
\end{itemize}

In summary we have explored some implications of new flavor interactions
at a scale a few orders of magnitude larger than the weak scale. When the
broken flavor gauge interaction is strong it can be expected to generate a
diverse set of multi-fermion operators in the low energy theory. We have
highlighted the role of mixed quark-lepton operators in the generation of
quark and lepton masses. A superweak theory of CP violation emerges very
naturally, in a manner of some relevance to the strong CP problem. In this
picture the smallness of CP violation in the quark sector and the smallness
of neutrino masses are related, since they both arise from effective
6-fermion operators.

\section*{Acknowledgement} This research was supported in part by the
Natural Sciences and Engineering Research Council of Canada.

\end{document}